\documentclass[twocolumn,pra,showpacs]{revtex4-1}
\usepackage{graphicx}
\usepackage{amssymb}
\usepackage{amsmath}
\usepackage{epsfig}
\usepackage{color}
\usepackage{mathtools}
\usepackage[colorlinks,linkcolor=blue,anchorcolor=blue,citecolor=blue,urlcolor=blue,dvipdfm]{hyperref}

\begin{document}

\title{Reviving the precision of multiple entangled probes in an open
system by simple $\pi $-pulse sequences}
\author{Yang Dong}
\author{Xiang-Dong Chen}
\author{Guang-Can Guo}
\author{Fang-Wen Sun}
\email{fwsun@ustc.edu.cn}
\affiliation{Key Lab of Quantum Information, Chinese Academy of Sciences,
School of physics, University of Science and Technology of China, Hefei,
230026, P.R. China, \\and Synergetic Innovation Center of Quantum Information $\&$ Quantum Physics, University of Science
and Technology of China, Hefei, 230026, P.R. China}
\date{\today}

\begin{abstract}
Quantum metrology with entangled states in realistic noisy environments always suffers from decoherence. Therefore, the measurement precision is greatly reduced. Here we applied the dynamical decoupling method to protect the $N$-qubit quantum metrology protocol and successfully revived the scaling of the measurement precision as ${N^{-k}}$ with $k\in \left[{5/6,11/12}\right]$. The degree of the precision revival, as determined by the noise spectrum distribution, indicates that the performance of the protected protocol can be further improved by controlling the noise spectrum. Such a protected protocol is proved to be universal for entanglement-based quantum metrology in the pure dephasing and relaxation noise, which should stimulate the development of practical quantum metrology for weak signal detection of microscopic physics.
\end{abstract}
\pacs{42.50.Lc, 03.65.Yz, 06.20.-f}
\maketitle
\section{Introduction}

Improving the resolution of spectroscopy is the heart of metrology. It is also of importance for science and technology. Over the past few decades, improvements with quantum resources \cite%
{3PhysRevLett112150802,3PhysRevLett112150801,3Giovannetti19112004,3PhysRevLett96010401,3giovannetti2011advances,3PhysRevLett110153901}
have been widely explored. Among these, the entanglement-based spectroscopy
\cite{3PhysRevD231693,3PhysRevA46R6797,3sun2008experimental} is an impressive method. It has been
shown that $N$-particle maximally entangled states in fully coherent
evolution (FCE) can be used to achieve the Heisenberg quantum limit (HQL),
in which the uncertainty can in principle scale as ${%
N^{-1}}$. However, the maximally entangled states are fragile in realistic noisy environments and the
notorious quantum decoherence would reduce the precision to
the standard quantum limit (SQL) scaling of ${%
N^{-1/2}}$ \cite{3PhysRevLett793865} in the detection
process. To keep the entanglement-based method alive, it is
necessary to avoid or fight against the decoherence. Recent studies
have shown that a super-classical scaling relationship can survive \cite%
{3PhysRevLett111120401} by avoiding the spatial direction of Markovian
noise. Generally, it is difficult to address Markovian noise and draw the
coherence from the environment \cite{3PhysRevLett793865}. However, in
solid spin systems \cite%
{3Childress13102006,3PhysRevLett103220802,3PhysRevA75042305,3bechtold2015three}%
, the Markovian approximation treatment of the environment is not always valid
\cite{3PhysRevLett109233601}. Recently, the ${N^{-3/4}}$ scaling of uncertainty of detection a physical
parameter has been demonstrated in the
non-Markovian dephasing environments \cite%
{3PhysRevLett109233601,3PhysRevA84012103,3PhysRevLett115170801}. Therefore,
the optimal measurement precision with maximally entangled states can be
improved by beating the non-Markovian noise of the sensor system, thus providing
an effective method to revive the HQL in realistic noisy environments.\par

\begin{figure}[tbp]
\centering
\includegraphics[width=8cm]{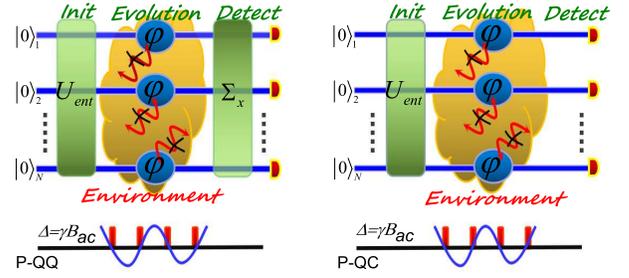}
\caption{Two protected quantum metrology protocols with N sensor qubits. The CPMG sequences (red rectangles) are applied
on each sensor qubit during the detection of a weak AC
magnetic field to protect sensor qubits.}
\label{3probescheme}
\end{figure}

In the past few decades, the dynamical decoupling (DD) method has been shown to be highly effective for suppressing decoherence by
filtering of the noise spectrum \cite%
{3PhysRevB77174509,313672630108083024,3PhysRevLett98100504,3zhong2015optically,3bar2012suppression,3PhysRevLett101180403,3PhysRevB85115303,Sekatski}%
. It has been well applied in one-qubit and two-qubit quantum information processes \cite{3PhysRevLett112050502,3PhysRevLett115110502}. Here we combine the traditional quantum metrology strategies, the quantum-classical (QC) and quantum-quantum (QQ) metrology protocol \cite{3PhysRevLett96010401,3giovannetti2011advances}, with the DD method to construct protected QC (P-QC) and protected QQ (P-QQ) protocols as shown in Fig. 1. The new protected protocols are robust against pure dephasing and longitudinal relaxation noise and can be used to revive the measurement precision scaling of ${N^{-k}}$ with $k\in \left[{5/6,11/12}\right]$. It is much better than the presented in Refs. \cite%
{3PhysRevLett109233601,3PhysRevA84012103,3PhysRevLett115170801}. The index $k$, which evaluates the degree of revival of the measurement precision by the DD protection method, is determined by the shape of the noise spectrum, especially its high frequency component. Consequently, in addition to extending the system coherence time, the performance of quantum metrology can be highly improved by changing the shape of the system noise spectrum. Furthermore, by studying a general quantum metrology protocol, the DD protection method is proved to be universal for reviving the advantage of the entanglement-based quantum metrology in pure dephasing and longitudinal relaxation noisy environments. A hybrid system composed of a superconducting circuit and $N$ $N{V^ - }$ centers in bulk diamond can be applied to present the protected quantum metrology protocol with current techniques. Therefore, the P-QQ and P-QC metrology protocols will stimulate the development of practical quantum metrology application in realistic environments and open a new avenue for weak signal detection. \par

\section{INDEPENDENT PURE DEPHASING NOISE MODEL}

Here we consider a quantum metrology protocol in a practical decoherent environment that can be
described with a fully quantum independent spin-boson noise model \cite%
{3RevModPhys591,3PhysRevLett98100504,3PhysRevLett109233601,313672630108083024,3PhysRevLett101180403}%
. The sensor system consists of $N$ two-level qubits, where ${\left\vert
{0_{j}}\right\rangle }$ ( ${\left\vert {1_{j}}\right\rangle }$) denotes the ground (excited) state of the $j$th qubit with the eigenvalue of $-1$ ($1$). The interaction
on the sensor qubits is given by
\begin{equation}
\begin{aligned} H =& \sum\nolimits_{j = 1}^N {\frac{{{\sigma
_{z,j}}}}{2}\sum\nolimits_i {{\lambda _{i,j}}(b_{i,j}^\dag + {b_{i,j}})} }
\\&+ \sum\nolimits_{j = 1}^N {\sum\nolimits_i {{\omega _{i,j}}b_{i,j}^\dag
{b_{i,j}}} } \text{,} \end{aligned}
\end{equation}%
where ${{\lambda _{i,j}}}$ denotes the coupling strength between the $j$th sensor
qubit and the environment. ${\sigma _{x,j}},{\sigma _{y,j}},{\sigma _{z,j}}$ are the
$j$th sensor qubit Pauli matrices. Here, the
environment is regarded as a bosonic bath with the annihilation (creation)
operator ${{b_{i,j}}}$ (${b_{i,j}^{\dag }}$) and frequency ${{\omega _{i,j}%
}}$ interacting with the $j$th sensor qubit. The relevant spin-boson bath property of the $j$th
sensor qubit is controlled by the completely high-energy cutoff noise model \cite{3RevModPhys591}:
\begin{equation}
\begin{aligned} {J_j}(\omega ) &= \sum\nolimits_i {{{\left| {{\lambda
_{i,j}}} \right|}^2}\delta (\omega - {\omega _{i,j}})} \\ &= 2{\alpha
_j}\omega \Theta ({\omega _{D,j}} - \omega ), \end{aligned}
\end{equation}%
where ${\alpha _{j}}$ is the dimensionless coupling strength and ${\omega _{D,j}}$ is the cutoff frequency. Without loss of generality, we discuss the quantum
metrology for the detection of an AC magnetic field $B$. The sensor state will gain the
shift of $\Delta ={\gamma }B$ after the interaction with the magnetic field,
where ${\gamma }$ is the gyromagnetic ratio of the sensor spin. Therefore, when the
sensor system is in this field, the evolution process can be described by
\begin{equation}
{H_{s}}=\Delta \cos \omega t\sum\nolimits_{j=1}^{N}{{\sigma _{z,j}}}+H.
\end{equation}

\section{THE PROTECTED QUANTUM METROLOGY PROTOCOL}

Because the final measurement is the only difference between the P-QC and P-QQ metrology protocols, we first discuss the P-QQ metrology protocol. The protected quantum metrology protocol can be divided into three typical steps \cite%
{3PhysRevLett96010401,3giovannetti2011advances}:\par

i) Initialization of the sensor system. Generally, quantum metrology can be
well realized with the Greenberger-Horns-Zeilinger ($\left\vert {GHZ}%
\right\rangle $) state described by
\begin{equation}
\left\vert {GHZ}\right\rangle ={D_{y}}(\pi /2)\left\vert G\right\rangle =%
\left[ {\left\vert G\right\rangle +\left\vert E\right\rangle }\right] /\sqrt{%
2}\text{,}
\end{equation}%
where ${D_{y}}(\pi /2)=\exp (-i\pi {\Sigma _{y}}/4)$ and ${\Sigma _{y}}%
=-i\left\vert E\right\rangle \langle G|+i\left\vert G\right\rangle \langle
E| $ with $\left\vert G\right\rangle ={\Pi }_{j=1}^{N}{\left\vert {0_{j}}%
\right\rangle }$ ($\left\vert E\right\rangle ={\Pi }_{j=1}^{N}{\left\vert 1{%
_{j}}\right\rangle }$).\par

ii) Evolution in the physical field to be detected. The sensor state will
gain a relative phase shift after interaction with the field for time $%
\tau $. In the protected measurement process, the evolution is accompanied by
symmetric timing dynamical decoupling Carr-Purcell-Meiboom-Gill (CPMG)
sequences \cite{3PhysRevA85032306}, which can be described by the operation ${%
\Pi _{x,N}}={\Pi }_{j=1}^{N}{i{\sigma _{x,j}}}$ acting on the sensor system for $n$ times,
as shown in Fig. 1. When the AC field picks up a negative sign at ${\delta
_{m}}=\frac{{m-1/2}}{n}\tau $ with $m=1,...n$, the $\pi $ pulse appears and flips all the
sensor qubits to generate an additive phase and dynamically decouple from fluctuating noisy bath. Therefore, the evolution process can be
represented as
\begin{equation}
\left\vert \Psi \right\rangle =R\left\vert {GHZ}\right\rangle \text{,}
\end{equation}%
where
\begin{align}
R=& \mathrm{e}{^{-i{H_{s}}({\delta _{n+1}}-{\delta _{n}})}}{\Pi _{x,N}}%
\mathrm{e}{^{-i{H_{s}}({\delta _{n}}-{\delta _{n-1}})}}{\Pi _{x,N}}\cdots
\notag \\
& \cdots \mathrm{e}{^{-i{H_{s}}({\delta _{2}}-{\delta _{1}})}}{\Pi _{x,N}%
\mathrm{e}}^{-i{H_{s}}({\delta _{1}}-{\delta _{0}})}\text{.}
\end{align}\par

iii) Detection process. After the detection with ${\Sigma _{x}}=\left\vert
E\right\rangle \langle G|+\left\vert G\right\rangle \langle E|$, the signal
will be
\begin{equation}
{s_{n}}(\tau )=\langle G|D_{y}^{\dag }(\mathrm{\pi }/2){R^{\dag }}{\Sigma
_{x}}R{D_{y}}(\mathrm{\pi }/2)\left\vert G\right\rangle \text{.}
\end{equation}

The shift $\Delta $ can be estimated by repeating the above measurement
processes $l = {T_t}/\tau $ times. Here, ${T_t} $ is the total duration and $\tau $ is the interrogation time of the experiment. The final signal can be directly calculated:
\begin{equation}
{s_{n}}(\tau )=\cos {N\varphi}\exp \left( {-2{\chi _{n}}}\right) ,
\end{equation}%
where
\begin{equation*}
{\varphi}=\frac{{4\Delta \tau }}{{\pi }}\text{,}
\end{equation*}%
\begin{equation*}
{\chi _{n}}=\sum\nolimits_{j=1}^{N}{\int\nolimits_{0}^{+\infty }{\frac{{{%
F_{n}}(\omega \tau ){J_{j}}(\omega )}}{{4{\omega ^{2}}}}}\coth (\omega /2{%
k_{B}}T)\mathrm{d}\omega }\text{,}
\end{equation*}%
\begin{equation*}
{F_{n}}(\omega \tau )=8{\sin ^{4}}\frac{{\omega \tau }}{{4n}}{\sin ^{2}}%
\frac{{\omega \tau }}{2}/{\cos ^{2}}\frac{{\omega \tau }}{{2n}}\text{.}
\end{equation*}%
Here ${k_{B}}$ is the Boltzmann constant and $T$ is the temperature of the environment. The behavior of ${\chi _{n}}$, which is the result of the
independent noise model for the most realistic solid system, governs the
detection precision. The phase is linearly related to the number of sensor
qubits $N$ and interrogation time $\tau $. Simply, $n=1$ corresponds to the spin
echo sequence with ${\chi _{1}}=\sum\nolimits_{j=1}^{N}{\int\nolimits_{0}^{+%
\infty }{\frac{{{F_{1}}(\omega \tau ){J_{j}}(\omega )}}{{4{\omega ^{2}}}}}%
\mathrm{d}\omega }$ and ${F_{1}}(\omega \tau )=8\sin {^{4}}\frac{{\omega
\tau }}{4}$. When the measurement operator ${\Sigma _{x}}$ is replaced with $%
{(-i)^{N}}{\Pi _{x,N}}/2$, the P-QQ metrology protocol is transferred to the P-QC metrology protocol as
shown in Fig. 1. In this case, the above conclusions are also valid.\par

\section{RESULTS OF THE PROTECTED QUANTUM METROLOGY PROTOCOL WITH CPMG SEQUENCES}

For $T\rightarrow 0K$, corresponding to the low-temperature limit, ${\chi _{n}}=\sum\nolimits_{j=1}^{N}{\int\nolimits_{0}^{+%
\infty }{\frac{{{F_{n}}(\omega \tau ){J_{j}}(\omega )}}{{4{\omega ^{2}}}}}%
\mathrm{d}\omega }=N\int\nolimits_{0}^{+\infty }{\frac{{{F_{n}}(\omega \tau )%
\bar{J}(\omega )}}{{4{\omega ^{2}}}}}\mathrm{d}\omega ,$ where ${\bar{J}%
(\omega )}$ is the equivalent noise spectral function. When ${\omega _{D}}%
\tau <<1$, the decoherence process can be characterized by
\begin{equation}
{\chi _{n}}\approx \alpha N{\tau ^{6}}
\end{equation}%
with $\alpha =\int\nolimits_{0}^{+\infty }{\frac{{{\omega ^{4}}\bar{J}%
(\omega )}}{{2\cdot {{(4n)}^{4}}}}}d\omega $. The precision enhancement over
SQL with entangled probe \cite{3PhysRevLett109233601} is $r={N^{(v-1)/2v}}$
where $v$ is the power law index of $\tau $. Therefore, for the present
protected protocol with the $\left\vert {GHZ}%
\right\rangle $ state, the enhancement is $r={N^{5/12}}$ and the final
precision is $\delta {\Delta _n} \sim {N^{ - 11/12}}$, very close to the HQL as shown in Fig. 2.
Even for the simplest spin-echo sequence, the optimal precision of entangled
probes follows the characteristic scale as ${N^{-7/8}}$, which is superior to ${N^{-3/4}}$ scaling obtained for the non-Markovian noise without DD protection \cite%
{3PhysRevLett109233601,3PhysRevA84012103,3PhysRevLett115170801}. The
optimal interrogation time of protected protocol is reasonable: $\tau
\approx 0.1/{\left( {12\alpha N}\right) ^{1/6}}$ or AC signals with
frequencies of approximately $10\mathrm{kHz}\thicksim 100\mathrm{kHz}$ after considering the
solid spin sensor system \cite%
{300344885775056503,3PhysRevLett115170801,3PhysRevX5041001}.\par

\begin{figure}[tbp]
\centering
\includegraphics[width=8cm]{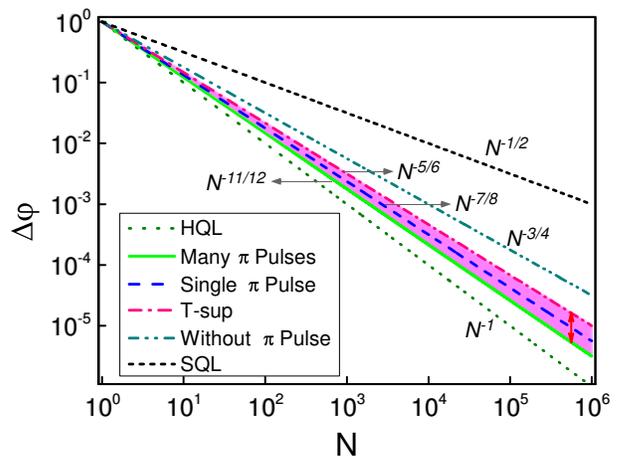}
\caption{Scaling of the measurement uncertainty versus the
number of sensor qubits. The SQL and HQL scalings are shown with black short-dashed and olive dotted lines, respectively. The dark cyan dashed dot-dotted, blue dashed and green solid lines represent the scalings of ${\protect\delta %
\Delta }$ with no pulse, a single $\protect\pi $ pulse and many $\protect\pi$ pulses, respectively, at low-temperature with a high-energy cutoff noise model. The region marked by the rad arrow indicates the scales ${\protect\delta \Delta }$ after taking temperature into consideration for many $\protect\pi $ pulses under non-Markovian dephasing.}
\label{3NOONresult}
\end{figure}

\section{HIGH-TEMPERATURE LIMIT AND THE LONGITUDINAL SPIN RELAXATION}

However, the high-energy cutoff noise model is too simple to describe the entire physical noise behaviors and the condition $T\rightarrow 0\mathrm{K}$ is always difficult to be achieved for a realistic sensor system, especially in the solid spin systems
surrounded by magnetic nuclei. When $T\rightarrow \mathrm{mK}$, the high-temperature approximation ( ${k_{B}}T>>{\gamma _{n}}B$) is always valid for
typical experimental conditions \cite{3Shi06032015} with ${\gamma _{n}}\sim
\mathrm{kHz/G}$ and $B\sim 100\mathrm{G}$, where ${\gamma _{n}}$ is the gyromagnetic ratio of the proton. Therefore, in these realistic systems, the decoherence of the sensor system is governed by thermal fluctuations and the system exhibits classical behavior \cite{3PhysRevLett98100504}. The effect of the environment can be
described by the classical Hamiltonian
\begin{equation}
H=\sum\nolimits_{j=1}^{N}{{f_{j}}(t){\sigma _{z,j}}}/2,
\end{equation}%
where ${{f_{j}}(t)}$ is a classical random noise field distribution on the $j$th
sensor qubit with $\left\langle {{f_{j}}(t)}\right\rangle =0$ and $%
\left\langle {{f_{j}}({t_{1}}){f_{j}}({t_{2}})}\right\rangle ={g_{j}}({t_{1}}%
-{t_{2}})$. After application of the protected metrology protocol, we can obtain a
similar result
\begin{equation}
\lim\limits_{T\rightarrow +\infty }{s_{n}}(\tau )=\cos {N\varphi}\exp
\left( {-2\lim\limits_{T\rightarrow +\infty }{\chi _{n}}}\right)
\end{equation}%
with
\begin{equation}
\lim\limits_{T\rightarrow +\infty }{\chi _{n}}=\sum\nolimits_{j=1}^{N}{%
\int\nolimits_{0}^{+\infty }{\frac{{{F_{n}}(\omega \tau ){p_{j}}(\omega )}}{{%
\pi {\omega ^{2}}}}}\mathrm{d}\omega }\text{,}
\end{equation}%
where $4{p_{j}}(\omega )/\pi $ is the classical noise power spectrum corresponding to the Fourier transform of ${g_{j}}(t)$. Here we consider two spin bath models with Lorentzian and Gaussian noise spectral density, respectively. The results are presented in Table 1.
\begin{table}[h]
\centering
\caption{Results for two common classical noise models with $\tau  < {\tau _c}$. The subscript is the number of $\pi$ pulses applied on each sensor qubit.}
\tabcolsep0.06in
\begin{tabular}{ccccccc}
\hline
\hline
& $\chi _{0}$ & $\delta \Delta _{0}$ & $\chi _{1}$ & $\delta \Delta _{1}$ & $\chi _{2}$ & $\delta \Delta _{2}$ \\ 
Lorentzian & $N\tau ^{2}$ & $N^{-3/4}$ & $N\tau ^{3}$ & $N^{-5/6}$ & $N\tau ^{3}$ & $N^{-5/6}$ \\ 
Gaussian & $N\tau ^{2}$ & $N^{-3/4}$ & $N\tau ^{4}$ & $N^{-7/8}$ & $N\tau ^{6}$ & $N^{-11/12}$ \\ \hline\hline
\end{tabular}
\end{table}

When the correlation time ${\tau _c}$ of the noisy environment is longer than the interrogation time ${\tau}$, the present protection protocol can also beat the SQL again in the common realistic experimental environment as shown in the Table 1. Additionally, the degree of reviving of the measurement precision by the DD sequences is determined by the shape of the noise spectrum, especially its high frequency component. The filter function of DD sequences exhibits weaker performance in the high-frequency region. Therefore, the measurement precision under the Gaussian noise spectrum is better than under the Lorentzian noise spectrum because of the much smaller high-frequency component of the Gaussian noise spectrum. This indicates that, in addition to the decoherence time, the quantum metrology performance is also controlled by the shape of the system noise spectrum. Moreover, Uhrig-DD(UDD) was proved to be universal for the suppression of qubit longitudinal spin relaxation noise \cite{3PhysRevLett101180403}. Because the CPMG sequence and UDD sequence have an intersection: the spin echo sequence and two $\pi$ pulses sequence, using a similar derivation, it can be shown that our protected protocols are also valid for the longitudinal relaxation noise. However, ${p_{j}}(\omega )$, which changes with different sensor qubits and local magnetic field, is always too complicated to be described by a specific function in reality \cite{3136726301410103041,3liu2012controllable,3bechtold2015three}. Recent experiment showed that ${\chi _{n}}\approx \alpha {\tau ^{v}}$ with $v\in (3,6)$ for single qubit under DD sequence \cite{3PhysRevLett105200402}. Hence, it indicates the optimal resolution of entangled probe follow a characteristic scaling as  ${N^{-k}}$ with $k\in \left[ {5/6,11/12} \right] $ in high temperature, as shown in Fig. 2 with pink area.
This result is also better than the previously obtained results \cite%
{3PhysRevLett109233601,3PhysRevA84012103,3PhysRevLett115170801}, demonstrating the
advantages of the protected protocols.\par
\begin{figure}[tbp]
\centering
\includegraphics[width=8cm]{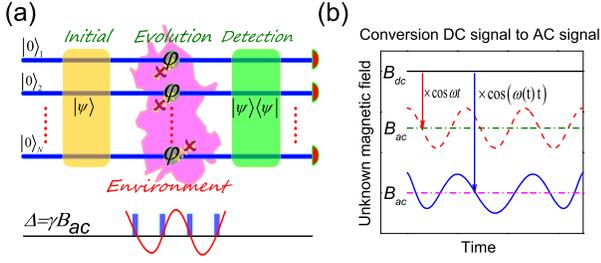}
\caption{(a) General protected quantum metrology protocol. Initial
sensor system in $\left\vert {\protect\psi }\right\rangle $ state is placed
into the magnetic field for time $\protect\tau $ accompanied by DD
sequences. The measurement operation is $\left\vert {\protect\psi }%
\right\rangle \left\langle {\protect\psi }\right\vert $. (b) The DC field is converted to arbitrary AC field by rotating the sensor system to match the DD sequences for the protected quantum metrology protocol well.}
\label{3GPMP}
\end{figure}
\section{GENERALIZATION OF PROTECTED QUANTUM METROLOGY PROTOCOL}

The above conclusion is drawn based on the detection of an AC
physical field with a specific initial state---$\left\vert {GHZ}%
\right\rangle $. Such a protected quantum metrology protocol can also be
generalized for wider practical applications \cite%
{3PhysRevLett96010401,3giovannetti2011advances,3PhysRevLett793865}.\par

i) Generalization of the sensor state. As
shown in Fig. 3(a), we consider a general quantum metrology
protocol with an initial state $\left\vert {\psi }\right\rangle $ and
detection operator $\left\vert {\psi }\right\rangle \left\langle {\psi }%
\right\vert $ that can be used to achieve the HQL in FCE. Therefore, the evolution generator \cite{3PhysRevLett96010401} $%
{h_{s,c}}=\sum\nolimits_{j=0}^{N}{{\sigma _{z,j}}}$ should satisfy the
relationship $\delta {h_{s,c}}\propto N$ to obtain the measurement precision scaling as $1/N$

\begin{equation}
\begin{aligned}
{\text{min}}\{ \delta {\Delta _{c,n}}\}&\leqslant {\left. {\delta {\Delta _{c,n}}} \right|_{\varphi  \to 2m\pi }} \\&\approx \sqrt {\frac{{{\pi ^2}}}{{16{\tau ^2}l\delta h_{s,c}^2}}} \propto 1/N \text{,}
\end{aligned}
\end{equation}
where $m$ is an integer. After considering the completely high-energy cutoff noise model, the result with the CPMG protection sequences is
\begin{equation}
\begin{aligned}
min\{ \delta {\Delta _n}\} &\leqslant {\left. \delta {\Delta
_n} \right|_{\scriptstyle N\varphi = 2m\pi + \theta,\theta \approx 0.01 \hfill
\atop {\scriptstyle \tau \approx 0.1/{(\alpha N)^{1/6}} \hfill }} }
\\& \approx {\left. {\sqrt {\frac{{{\theta ^2}\delta h_{s,c}^2/4 + 2\alpha N{\tau ^6}}}{{l{\theta ^2}{{\left| {d\varphi /d\Delta } \right|}^2}\delta h_{s,c}^4/4}}} } \right|_{\theta \approx 0.01 \hfill
\atop {\scriptstyle \tau \approx 0.1/{(\alpha N)^{1/6}} \hfill }} }
\\& \propto {N^{ - 11/12}}\text{.}
\end{aligned}
\end{equation}

This shows that the general protected quantum metrology protocol is universal, thus preserving the advantage of an arbitrary entangled state that can be used to beat the SQL in pure dephasing noisy environment. And it is also valid for longitudinal relaxation noise.  There exist some other entangled
states that possess the character of super-classical scaling relationship
in the FCE beyond the above quantum metrology protocol and are relatively easy to generate \cite%
{3PhysRevLett115170801}. So in these cases, the
protection with the DD method is always valid because it changes the decoherence process
when $\tau \approx 0.1/{(\alpha N)^{1/6}}$. For example, a spin cat state or
a two-axis twisted state can obtain $\delta {\Delta _n} \propto {N^{ - k}}$ with $k\in
\left[{5/6,11/12}\right] $.\par

ii) Generalization of detection frequency region. In the above discussion,
an AC field can be detected well with the protected protocol. For the
measurement of a DC field, a simple method is to rotate the sensor system in
the field and introduce an AC interaction. Hence, the DC physical field can be intentionally
transferred to an arbitrary AC field on purpose, as shown in Fig. 3(b). For
example, with the advances in laser-induced rotation technology, the rotation
speed of the solid sensor system, such as a negatively charged
nitrogen-vacancy center in diamond \cite{3neukirch2015multi}, can be
controlled with a circularly polarized trapping laser beam \cite%
{3padgett2011tweezers,3arita2013laser}. Such a rotation can be modified to
extend the detection frequency to the $\mathrm{MHz}$ \cite{3arita2013laser}.
Moreover, the controlled frequency may be fitted to the UDD sequence
\cite{3PhysRevLett98100504,313672630108083024} to optimally protect the
sensor state in noisy environments, further enhancing the
estimation resolution approaching the HQL.\par

\begin{figure}[tbp]
\centering
\includegraphics[width=8.9cm]{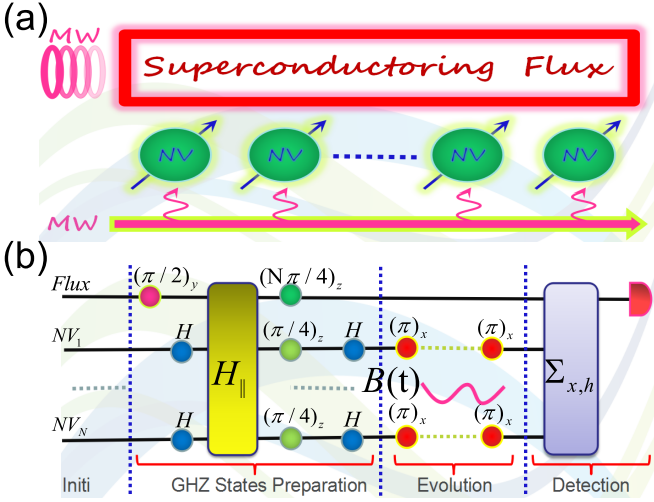}
\caption{(a) The schematic diagram of hybrid sensor system. All qubits can be operated by microwave pulse. (b) The quantum circuit for creating the  $N+1$-qubits GHZ state and detecting weak AC physical field. Here, ${(\theta )_b} = {e^{ - i\theta {{\hat \sigma }_b}/2}}$ and $b = x,y,z$. }
\label{3NOONsystemandcircuit}
\end{figure}

\section{CONSTRUCTION OF PROTECTED METROLOGY PROTOCOL WITH HYBRID SYSTEM}

As shown in Fig. 4(a), a hybrid system composed of a superconducting circuit and $N$ $N{V^ - }$ centers in bulk diamond can be used to present the protected quantum metrology protocol, where the superconducting flux qubit is a control and readout qubit and spin states in NV center are memory and sensor qubits. In the interaction picture, the Hamiltonian of the hybrid sensor system
can be written as \cite{3PhysRevLett115170801}:
\begin{equation}
\begin{aligned} {H_{total}} = {H_{||}} + {H_ \bot } \text{,} \end{aligned}
\end{equation}
with
\begin{equation}
\begin{aligned} {H_{||}} = \hbar {g_1}\hat \sigma _z^{(c)}\hat J_z^{(m)} \text{,}
\end{aligned}
\end{equation}
\begin{equation}
\begin{aligned} {H_ \bot } = \hbar {g_2}(\hat \sigma _ + ^{(c)}\hat J_ -
^{(m)} + \hat \sigma _ - ^{(c)}\hat J_ + ^{(m)})\text{,} \end{aligned}
\end{equation}
where ${H_{||}}$(${H_ \bot }$) is the longitudinal (transverse) interaction
between control qubit and sensor qubit, $%
\hat \sigma _z^{(c)}$ denotes Pauli operator acting on control qubit, $\hat
J_z^{(m)}$ is collective spin operator on sensor qubit, $\hat \sigma
_ \pm ^{(c)}$($\hat J_ \pm ^{(m)}$) is a ladder operator on
control (sensor) qubit. In order to briefly demonstrate the
original idea to show the P-QQ metrology protocol, we make use of longitudinal Hamiltonian \cite%
{3PhysRevLett793865} to create ($N+1$)-qubit $\left| {GHZ + } \right\rangle = %
\left[ {{{\left| G \right\rangle }_h} + {{\left| E \right\rangle }_h}} %
\right]/\sqrt 2 $ ,where ${\left| G \right\rangle _h} = {\left| 0
\right\rangle _f}\prod\nolimits_{j = 1}^N {{{\left| 0 \right\rangle }_{NV_j^
- }}} $ and ${\left| E \right\rangle _h} = {\left| 1 \right\rangle _f}%
\prod\nolimits_{j = 1}^N {{{\left| 1 \right\rangle }_{NV_j^ - }}} $. The single-qubit logic gate can
be made with microwave operation and the two-qubit Controlled-NOT gate, or
CNOT, can be directly constructed with
\begin{equation}
\begin{aligned} {U_{CNOT}} = \left( {I \otimes H} \right){U_{CZ}}\left( {I
\otimes H} \right) \text{,} \end{aligned}
\end{equation}
and
\begin{equation}
\begin{aligned} {U_{CZ}} = \sqrt i {e^{ - i\frac{\pi }{4}\sigma _z^c}}{e^{ -
i\frac{\pi }{4}\sigma _z^t}}{e^{ - i\frac{{3\pi }}{4}\sigma _z^c\sigma
_z^t}}\text{,} \end{aligned}
\end{equation}
where $I$, $H$ and ${U_{CZ}}$ are identity, Hadamard, and Controlled-Z gates, respectively. $\sigma _z^c$ ($\sigma _z^t$) denotes pauli operator acting on the
control (target) qubit. Hence, the ($N+1$)-qubit CNOT gate can be created:
\begin{equation}
\begin{aligned} \prod\limits_{j = 1}^N {{C_f}NO{T_{NV_j^ - }}} =
\prod\limits_{j = 1}^N {{H_j}} \prod\limits_{j = 1}^N {{U_{CZ,j}}}
\prod\limits_{j = 1}^N {{H_j}} \text{,} \end{aligned}
\end{equation}
\begin{equation}
\begin{aligned} \prod\limits_{j = 1}^N {{U_{CZ,j}}} = \prod\limits_{j = 1}^N
{\sqrt {{i^N}} {e^{ - i\frac{{N\pi }}{4}\hat \sigma _z^{(c)}}}{e^{ -
i\frac{\pi }{4}\hat J_z^N}}{e^{ - i\frac{{3\pi }}{{4{g_1}}}{H_{||}}}}}\text{,}
\end{aligned}
\end{equation}
where ${{C_f}NO{T_{NV_j^ - }}}$ denotes a CNOT gate which flux is control
qubit and ${NV_j^ - }$ is target qubit, ${H_j}$ is a Hadamard gate acting on
${NV_j^ - }$, and ${{U_{CZ,j}}}$ is Controlled-Z gate acting on flux and ${%
NV_j^ - }$. Omitting the global phase $\sqrt {{i^N}} $, we can get ($N+1$)-qubit ${%
\left| {GHZ+} \right\rangle }$ state and those processes can be denoted by an
operations ${U_{ent,h}}$. If DD sequences \cite%
{3PhysRevLett112050502,3PhysRevLett115110502} are interleaved in the generation processes, the decoherence effects can
be neglected. It also holds for detection process. Then the hybrid system
evolves in the weak AC physical field for time $\tau $ with
symmetric timing dynamical decoupling CPMG sequence on each sensor
qubit. Finally, the measurement operation can be constructed in this way:
\begin{equation}
\begin{aligned} {\Sigma _{x,h}} &= {\left| G \right\rangle _{hh}}\langle E|
+ {\left| E \right\rangle _{hh}}\langle G| \\&= \frac{{\left| {GHZ + }
\right\rangle \langle GHZ + |}}{2} - \frac{{\left| {GHZ - } \right\rangle
\langle GHZ - |}}{2} \text{,}\end{aligned}
\end{equation}
with $\left| {GHZ - } \right\rangle = \left[ {\left| G \right\rangle -
\left| E \right\rangle } \right]/\sqrt 2 $. The former measurement
operator is created by ${\left| {GHZ+} \right\rangle \langle GHZ+|}={%
U_{ent,h}}{\left| G \right\rangle _{hh}}\langle G|U_{ent,h}^{ - 1}.$ By just
adding a ${(\pi )_z}$ gate on flux qubit following closely after $U_{ent,h}$%
, the measurement operator of $\left| {GHZ - } \right\rangle \left\langle {%
GHZ - } \right|$ is created and the measurement operation ${\Sigma _x}
$ can be obtained. So briefly, the P-QQ metrology protocol can be put
into practice with current system and technology with simple few
manipulations. It also holds for P-QC and the general protected quantum metrology protocol.\par

\section{CONCLUSION}

By fighting the pure dephasing and longitudinal relaxation noisy environment with the DD method on an $N$-qubit quantum metrology protocol, we successfully revive the measurement precision
of the $\left\vert {GHZ}\right\rangle $ state scaling as ${N^{-11/12}}$ at low temperatures. Even in the high-temperature region, the precision can still be improved to ${N^{-k}}$ with $%
k\in \left[ {5/6,11/12}\right] $.
The degree of revival of the measurement precision with the DD protection is primarily determined by the high frequency component
of the noise spectrum, indicating that the performance of quantum metrology in realistic environments is highly dependent on the details of the noise spectrum in addition to the coherence time. Moreover, we generalize our protocol and prove that it can maintain the validity of the entanglement-based method in pure dephasing and longitudinal relaxation noisy environments. For experimental realization, the use of hybrid quantum circuits based on superconducting circuits interacting with quantum solid systems is very promising with current quantum techniques. Therefore, such a protected quantum metrology would show its high practical potential for the detection of ultra-weak physical parameters in realistic noisy environments.\par

\section*{Acknowledgment}
We thank Hailin Wang for fruitful discussions. This work was supported by the Strategic Priority Program(B) of the Chinese
Academy of Sciences (No. XDB01030200), the National Natural Science Foundation
of China (Nos. 11374290, 61522508, 91536219, 11504363), the Fundamental
Research Funds for the Central Universities and the China Postdoctoral Science Foundation (No.2015M571935).

\appendix
\section{The calculation of signal
expression}

The signal ${s_{n}}(\tau )=\langle G|D_{y}^{\dag }(\mathrm{\pi }/2){R^{\dag }}{\Sigma_{x}}R{D_{y}}(\mathrm{\pi }/2)\left\vert G\right\rangle$ can be calculated with the unitary transformation
\begin{equation}
\begin{aligned} U = \exp \left( {\sum\nolimits_{j = 1}^N {{\sigma
_{z,j}}{K_j}} } \right)\text{,} \end{aligned}
\end{equation}%
with ${K_j} = \sum\nolimits_{i = 1} {\frac{{{\lambda _{i,j}}}}{{2{\omega _{i,j}}}}(b_{i,j}^\dag  - {b_{i,j}})} .$ Under this transformation, ${H_{s}}$ can be diagonal
\begin{equation}
\begin{aligned} H_s^{ef} = {H^{ef}} + \Delta \cos \omega t\sum\nolimits_{j = 1}^N {{\sigma _{z,j}}}  \text{,} \end{aligned}
\end{equation}%
with ${H^{ef}} = \sum\nolimits_{j = 1}^N {\sum\nolimits_i {{\omega _{i,j}}b_{i,j}^\dag {b_{i,j}}} }  + E - \sum\nolimits_{j = 1}^N {\int_0^{ + \infty } {\frac{{{J_j}(\omega )}}{\omega }d\omega } } $. Here, $E$ denotes the energy offset of sensor system. Therefore, arbitrary quantum operator $F$ has a new form: ${F^{ef}}=UF{U^{\dag }}$. So the signal can be expressed as
\begin{equation}
\begin{aligned} {s_n}(\tau ) = \langle G|D_y^{ef\dag }(\pi /2){R^{ef\dag
}}\Sigma _x^{ef}{R^{ef}}D_y^{ef}(\pi /2)\left| G \right\rangle .
\end{aligned}
\end{equation}%
Since ${H_{s}}$ and even number $\pi $-pluses do not cause spin flip, we can
directly get
\begin{equation}
\begin{aligned} {s_n}(\tau ) &=  - \frac{i}{2}\langle G|\Sigma _y^{ef}{R^{ef\dag }}\Sigma _x^{ef}{R^{ef}}\left| G \right\rangle \\& + \frac{i}{2}\langle G|{R^{ef\dag }}\Sigma _x^{ef}{R^{ef}}\Sigma _y^{ef}\left| G \right\rangle
\\&= \operatorname{Im} \langle G|\Sigma _y^{ef}{R^{ef\dag }}\Sigma _x^{ef}{R^{ef}}\left| G \right\rangle \text{,}
\end{aligned}
\end{equation}%
with $\langle G|\Sigma _x^{ef}\left| G \right\rangle  = \langle E|\Sigma _x^{ef}\left| E \right\rangle  = \langle G|\Sigma _y^{ef}\left| G \right\rangle  = \langle E|\Sigma _y^{ef}\left| E \right\rangle  = 0$. \par
By defining the time-dependent operators
\begin{equation}
\begin{aligned} {F^{ef}}(\tau ) = \exp (i{H^{ef}}\tau ){F^{ef}}\exp ( -
i{H^{ef}}\tau )\text{,} \end{aligned}
\end{equation}%
the signal can be expressed as
\begin{equation}
\begin{aligned}
{s_n}(\tau ) = \operatorname{Im} \left[ {{e^{ - i{\varphi _n}}}\langle G|\Sigma _y^{ef}(0){{\tilde R}^\dag }\Sigma _x^{ef}(\tau )\tilde R\left| G \right\rangle } \right]
\end{aligned}
\end{equation}
with
\begin{equation}
\begin{aligned}
{\varphi _n} &= \sum\nolimits_{i = 0}^n {\int_{{\delta _i}}^{{\delta _{i + 1}}} {\Delta \cos \omega tdt} }
\\& = \frac{{4\Delta N\tau }}{\pi }\text{,}
\end{aligned}
\end{equation}
\begin{equation}
\begin{aligned}
\tilde R = \Pi _{x,N}^{ef}({\delta _n})\Pi _{x,N}^{ef}({\delta _{n - 1}}) \cdots \Pi _{x,N}^{ef}({\delta _1}).
\end{aligned}
\end{equation}
Similar relations \cite{3PhysRevLett98100504} are also held
\begin{equation}
\begin{aligned}
\Pi _{x,N}^{ef}({\delta _i})\left| {G/E} \right\rangle  = {i^N}\exp \left( { \pm 2\sum\nolimits_{j = 1}^N {{K_j}({\delta _i})} } \right)\left| {E/G} \right\rangle {\text{,}}
\end{aligned}
\end{equation}%
\begin{equation}
\begin{aligned} \Sigma _x^{ef}(\tau )\left| {G/E} \right\rangle = \exp
\left( { \pm 2\sum\nolimits_{j = 1}^N {{K_j}(\tau )} } \right)\left| {E/G}
\right\rangle \end{aligned} \text{,}
\end{equation}%
\begin{equation}
\begin{aligned} \Sigma _y^{{\text{ef}}}(\tau )\left| {G/E} \right\rangle =
\mp i\exp \left( { \pm 2\sum\nolimits_{j = 1}^N {{K_j}(\tau )} }
\right)\left| {E/G} \right\rangle. \end{aligned}
\end{equation}%
Here we can write the signal as:
\begin{equation}
\begin{aligned}
{s_n}(\tau ) = \operatorname{Im} \left[ {i{e^{ - i{N\varphi}}}\left\langle {{\operatorname{e} ^{2{\Lambda  _n}K}}} \right\rangle } \right] \text{,}
\end{aligned}
\end{equation}%
where
\begin{equation}
\begin{aligned}
{e^{2{\Lambda _n}K}} =&{\operatorname{e} ^{ - \sum\nolimits_{j = 1}^N {{K_j}(0)} }}{\operatorname{e} ^{2\sum\nolimits_{j = 1}^N {{K_j}({\delta _1})} }}{\operatorname{e} ^{ - 2\sum\nolimits_{j = 1}^N {{K_j}({\delta _2})} }} \cdots
\\&{\operatorname{e} ^{{{( - )}^{n + 1}}2\sum\nolimits_{j = 1}^N {{K_j}({\delta _n})} }}{\operatorname{e} ^{{{( - )}^n}2\sum\nolimits_{j = 1}^N {{K_j}(\tau )} }}  \times
\\&{\operatorname{e} ^{{{( - )}^{n - 1}}2\sum\nolimits_{j = 1}^N {{K_j}({\delta _n})} }} \cdots {\operatorname{e} ^{ - 2\sum\nolimits_{j = 1}^N {{K_j}({\delta _2})} }} \times
\\&{\operatorname{e} ^{2\sum\nolimits_{j = 1}^N {{K_j}({\delta _1})} }}{\operatorname{e} ^{ - \sum\nolimits_{j = 1}^N {{K_j}(0)} }} \text{.}
\end{aligned}
\end{equation}%
Based on the Baker-Campbell-Hausdorff (BCH) formula and $\left\langle {{%
e^{A}}}\right\rangle ={e^{\left\langle {{A^{2}}}\right\rangle /2}}$ for
linear bosonic operator A \cite{3PhysRevLett98100504}, we get the results in main text
\begin{equation}
\begin{aligned}
{s_n}(\tau ){\text{ }} &= \operatorname{Re} \left[ {{e^{ - i{N\varphi}}}\left\langle {{e^{2{\Lambda _n}K}}} \right\rangle } \right]
\\& = \cos {\varphi _n}\exp \left( { - 2{\chi _n}} \right)\text{,}
\end{aligned}
\end{equation}%
with
\begin{equation}
\begin{aligned}
{\Lambda _n}K &= {( - )^{n + 1}}\sum\nolimits_{j = 1}^N {[{{( - )}^n}{K_j}(0) - {K_j}(\tau)}  + 2\sum\nolimits_{i = 1}^n {{K_j}({\delta _i})} ]
\\& = - \sum\nolimits_{j = 1}^N {\sum\nolimits_i {\frac{{{\lambda _{i,j}}}}{{2{\omega _{i,j}}}}(b_{i,j}^\dag {y_n}({\omega _i}\tau ) - {b_{i,j}}y_n^*({\omega _i}\tau ))} }\text{,}
\end{aligned}
\end{equation}%
\begin{equation}
\begin{aligned}
{y_n}(z) = 1 + {( - )^{n + 1}}{e^{iz}} + 2\sum\nolimits_{i = 1}^n {{{( - )}^j}{e^{iz{\delta _j}/\tau }}}\text{.}
\end{aligned}
\end{equation}%
Hence, we have
\begin{equation}
\begin{aligned}
{\chi _n} &= \sum\nolimits_{j = 1}^N {\sum\nolimits_i {\frac{{\lambda _{i,j}^2}}{{4\omega _{i,j}^2}}{{\left| {{y_n}({\omega _i}\tau )} \right|}^2}\left\langle {b_{i,j}^\dag {b_{i,j}} + {b_{i,j}}b_{i,j}^\dag } \right\rangle } }
\\&= \sum\nolimits_{j = 1}^N {\int_0^{ + \infty } {\frac{{{F_n}(\omega \tau ){J_j}(\omega )}}{{4{\omega ^2}}}} \coth (\omega /2{k_B}T){\text{d}}\omega } \text{,}
\end{aligned}
\end{equation}%
where
\begin{equation}
\begin{aligned}
{{F_n}(\omega \tau ) = {{\left| {{y_n}(\omega \tau )} \right|}^2}}\text{,}
\end{aligned}
\end{equation}
\begin{equation}
\begin{aligned}
{J_j}(\omega ) &= \sum\nolimits_i {{{\left| {{\lambda _{i,j}}} \right|}^2}\delta (\omega  - {\omega _{i,j}})} \\ &= 2{\alpha _j}\omega \Theta ({\omega _{D,j}} - \omega )\text{.}
\end{aligned}
\end{equation}
When the measurement operator ${\Sigma _{x}}$ is
replaced with ${(-i)^{N}}{\Pi _{x,N}}/2$, we can transfer P-QQ metrology protocol to P-QC
metrology protocol. Due to similar relation ${( - i)^N}\Pi _{x,N}^{ef}(\tau )/2\left| {G/E} \right\rangle  = \exp \left( { \pm 2\sum\nolimits_{j = 1}^N {{K_j}(\tau )} } \right)\left| {E/G} \right\rangle $ is also held, the results of P-QC is the same as P-QQ.

\section{The proof of universality
for general metrology protocol}

Usually in the FCE, we describe the detection of a physical field by a
generator \cite{3PhysRevLett96010401}
\begin{equation}
\begin{aligned} {h_{s,c}} = \sum\nolimits_{j = 0}^N {{\sigma _{z,j}}}\text{.}
\end{aligned}
\end{equation}%
In quantum metrology, the fluctuation of $\delta {\Delta _{s,c}}=\delta {%
\varphi _{s,c}}/\tau $ is constrained by the generalized Heisenberg
uncertainty relation
\begin{equation}
\begin{aligned} \delta \varphi \delta {h_{s,c}}\geqslant1/(2\sqrt l )\text{,}
\end{aligned}
\end{equation}%
where ${\left\langle {\delta {h_{s,c}}}\right\rangle ^{2}}=\left\langle {%
h_{s,c}^{2}}\right\rangle -{\left\langle {{h_{s,c}}}\right\rangle ^{2}}$ is
the fluctuation of $h_{s,c}$ on the initial state \cite{3PhysRevLett96010401}%
. Hence, if we express the initial state as $\left\vert {\psi }\right\rangle
=\sum\nolimits_{x=0}^{{2^{N}}-1}{{a_{x}}\left\vert x\right\rangle }$ and $%
\sum\nolimits_{x=0}^{{2^{N}}-1}{{{\left\vert {{a_{x}}}\right\vert }^{2}}=1},$
we will get
\begin{equation}
\begin{aligned} \left\langle {{h_{s,c}}} \right\rangle =
2\sum\nolimits_{x=0}^{{2^N} - 1} {c(x)\left| a \right|_x^2} - N,
\end{aligned}
\end{equation}%
\begin{equation}
\begin{aligned} \left\langle {h_{s,c}^2} \right\rangle =
\sum\nolimits_{x=0}^{{2^N} - 1} {{{(2c(x) - N)}^2}\left| a \right|_x^2},
\end{aligned}
\end{equation}%
where the function of ${c(x)}$ counts number of "1" (excited state) in state $%
\left\vert x\right\rangle .$ The measurement uncertainty scaling
as $1/N$ is equal to
\begin{equation}
\begin{aligned}
{\left\langle {\delta {h_{s,c}}} \right\rangle ^2} &= \left\langle {h_{s,c}^2} \right\rangle  - {\left\langle {{h_{s,c}}} \right\rangle ^2}
\\& = 2\kappa
\\&  \propto {N^2}\text{,}
\end{aligned}
\end{equation}%
with
\begin{equation}
\begin{aligned} \kappa = \sum\nolimits_{x,y=0}^{{2^N} - 1} {{{(c(x) -
c(y))}^2}\left| a \right|_x^2\left| a \right|_y^2}\text{.} \end{aligned}
\end{equation}

So with the initial state $\left| {\psi} \right\rangle $ in the general
metrology protocol, the evolution, which is denoted by $R_c$, can be
described by Eq.(6) by dropping the noise term . The final result is:
\begin{equation}
\begin{aligned}
{p_{c,n}} &= \left\langle {{\psi }} \right|R_c^\dag \left|
{{\psi }} \right\rangle \left\langle {{\psi }} \right|{R_c}\left| {{\psi }}
\right\rangle
\\& = \sum\nolimits_{y = 0}^{{2^N} - 1} {{{\left| {{a_y}} \right|}^2}{e^{ic(y)\phi }}} \sum\nolimits_{x = 0}^{{2^N} - 1} {{{\left| {{a_x}} \right|}^2}{e^{ - ic(x))\phi }}}
\\&= \sum\nolimits_{x,y = 0}^{{2^N} - 1} {{{\left| {{a_x}}
\right|}^2}{{\left| {{a_y}} \right|}^2}{e^{i(c(y) - c(x))\phi }}}
\\&=\sum\nolimits_{x,y = 0}^{{2^N} - 1} {{{\left| {{a_x}} \right|}^2}{{\left| {{a_y}} \right|}^2}\sum\nolimits_{q = 0}^{ + \infty } {\frac{{{{\left[ {i(c(y) - c(x))\phi } \right]}^q}}}{{q!}}} }
\\& = \sum\nolimits_{x,y = 0}^{{2^N} - 1} {{{\left| {{a_x}} \right|}^2}{{\left| {{a_y}} \right|}^2}\sum\nolimits_{q = 0}^{ + \infty } {\frac{{{{\left[ {i(c(y) - c(x))\phi } \right]}^{2q}}}}{{(2q)!}}} }
\\& =\sum\nolimits_{x,y = 0}^{{2^N} - 1} {{{\left| {{a_x}} \right|}^2}{{\left| {{a_y}} \right|}^2}\cos (c(y) - c(x))\phi }\text{,}
\end{aligned}
\end{equation}
with $\phi = \frac{{4\Delta \tau }}{{\pi }}$. So the bound of frequency
uncertainty is given by
\begin{equation}
\begin{aligned} min\{ \delta {\Delta _{c,n}}\} &\leqslant{\left.
{\frac{{\sqrt {{p_{c,n}}(1 - {p_{c,n}})} }}{{\left| {d{p_{c,n}}/d\Delta }
\right|\sqrt l }}} \right|_{\phi \to 2m\pi }} \\&  \approx \sqrt {\frac{{{\pi ^2}}}{{32{\tau ^2}l\kappa }}}
\\& \propto 1/N \text{,} \end{aligned}
\end{equation}
where $m$ is an integer. The above relationship shows that the general
quantum metrology protocol is loyally to demonstrate any probe state
with the character scaling of $1/N$ in the FCE. After taking the completely high-
energy cutoff noise model into consideration, we have
\begin{equation}
\begin{aligned}
{p_n} & = \sum\nolimits_{x,y = 0}^{{2^N} - 1} {{{\left| {{a_x}} \right|}^2}{{\left| {{a_y}} \right|}^2}{e^{i(c(y) - c(x))\phi  - 2c(x \oplus y)\alpha {\tau ^6}}}}
\\&= \sum\nolimits_{x,y = 0}^{{2^N} - 1} {{{\left| {{a_x}} \right|}^2}{{\left| {{a_y}} \right|}^2}\cos \left[ {(c(y) - c(x))\phi } \right]{e^{ - 2c(x \oplus y)\alpha {\tau ^6}}}} \text{,}
\end{aligned}
\end{equation}
where $\oplus $ is a bitwise XOR operation and $\alpha$ is decoherence rate.
Hence, we can estimate the bound of frequency
uncertainty in realistic environment with DD protection
\begin{equation}
\begin{aligned} min\{ \delta {\Delta _n}\} &\leqslant {\left. {\frac{{\sqrt
{{p_n}(1 - {p_n})} }}{{\left| {d{p_n}/d\Delta } \right|\sqrt l }}}
\right|_{\scriptstyle N\phi = 2m\pi + \theta \hfill \atop {\scriptstyle \tau
\approx 0.1/{(\alpha N)^{1/6}} \hfill \atop \scriptstyle \theta \approx 0.1
\hfill}} } \\& \approx {\left. {\sqrt {\frac{{{\theta ^2}\kappa /2 + \xi }}{{l{\theta ^2}{\kappa ^2}{{\left| {d\varphi /d\Delta } \right|}^2}}}}}  \right|_{\scriptstyle\tau
\approx 0.1/{(\alpha N)^{1/6}}\hfill\atop \scriptstyle\theta \approx 0.1
\hfill}}\text{,}\\& \end{aligned}
\end{equation}
with $\xi = 2\alpha {\tau ^6}\sum\nolimits_{x,y}^{{2^N} - 1} {c(x \oplus
y)\left| a \right|_x^2\left| a \right|_y^2} \leqslant 2\alpha {\tau ^6}N.$
Therefore, we have
\begin{equation}
\begin{aligned} min\{ \delta {\Delta _{c,n}}\} &\leqslant {\left. {\sqrt {\frac{{{\theta ^2}\kappa /2 + 2\alpha N{\tau ^6}}}{{l{\theta ^2}{\kappa ^2}{{\left| {d\varphi /d\Delta  } \right|}^2}}}} }
\right|_{\scriptstyle\tau \approx 0.1/{(\alpha
N)^{1/6}}\hfill\atop \scriptstyle\theta \approx 0.1 \hfill}}
\\& \propto {N^{
- 11/12}} \text{.}\end{aligned}
\end{equation}
Generally, the condition $N\phi = 2m\pi + \theta $ and $\theta \approx 0.1
$ is applied to Taylor expansion $\cos \theta \approx 1 - {\theta
^2}/2$ legally. However, in the experiment with $\left\vert {GHZ}%
\right\rangle $ state, this condition can be replaced by $%
N\phi = \frac{{m_1\pi }}{2}$ with $m_1$ is an odd. For a superposition state with equal
probability amplitude, the condition should be $N\phi = 2{%
m_2}\pi + \frac{{{m_3}\pi }}{2}$ with $m_2$ and $m_3$ are integers which are much
smaller than $N$.

\bibliographystyle{apsrev}

\end{document}